\documentclass[12pt]{article}
\usepackage[pctex32]{graphics}
\begin{document}
\begin{center}
    
{\bf  Recombination of Intersecting D-branes in Tachyon Field Theory}

\vspace{1cm}

                      Wung-Hong Huang\\
                       Department of Physics\\
                       National Cheng Kung University\\
                       Tainan,70101,Taiwan\\

\end{center}
\vspace{3cm}

      The mechanism of recombination of intersecting D-branes is investigated within the framework of effective tachyon field theory.  We regard the branes as the kink-type tachyon condensed states and use the effective two-tachyon Lagrangian to study the off-diagonal fluctuations therein.   It is seen that there is a tachyonic mode in the off-diagonal fluctuations, which signs the instability of the intersecting D-branes.   We diagonalize the two-by-two tachyonic matrix field and  see that the corresponding eigenfunctions could be used to describe the new recombined branes.   Our prescription can be extended to discuss the general behavior of the recombination of  intersection D-branes with arbitrary shapes.   We present in detail the physical reasons behind the mathematical process of diagonalizing the tachyonic matrix field.

\vspace{4cm}
\begin{flushleft}
E-mail:  whhwung@mail.ncku.edu.tw\\
\end{flushleft}

\newpage
\section{Introduction}

    Branes and their different configurations are known to play important roles in the unified formulation of nonperturbative string/M theory [1-4].   Brane configurations that contain tachyonic model are unstable and will decay.  The decay models vary depending on the unstable brane configurations.   In a series of papers, Sen had made several conjectures on tachyon condensation [5] which  have drawn attention to various non-BPS D-brane configurations in string theory [6-10].  According to the conjectures the potential height of the tachyon potential exactly cancels the tension of the original unstable D-brane and, at the stable true vacuum, the original D-brane disappears and the kink-type tachyon condensed states correspond to lower-dimensional D-branes.  

    One of the interesting D-brane configurations is a pair of  D-branes intersecting  at an angle [11-14].  Such a scenario of intersecting branes is known to be possible  to construct models similar to the standard model [15] and may provide a simple mechanism for inflation in the early universe [16].   

   In this paper we will investigate the mechanism of recombination of intersection D-branes.  Within the framework of the effective tachyon field theory [6-10] we describe how a pair of D-branes intersecting at an angle shall  recombine.

      In the section II  we briefly review  the effective one-tachyon Lagrangian.   In the section III  we use the effective two-tachyon Lagrangian [9], in which the tachyon is described as a two-by-two matrix field, to study the diagonal fluctuation in the background of the kink solutions.  The kink solutions in here are regarded as tachyon condensations of the non-BPS brane in higher dimension.  As the tachyon condensated state represents a lower-dimensional  BPS-brane, it is a stable configuration.   Therefore the diagonal fluctuation in the background of the kink solutions will have no tachyonic mode.     Then, as the main contain of this paper, we study the off-diagonal fluctuation in the background of the kink solutions.  in this case as the D-branes (i.e. kinks) intersecting at an angle is unstable the off-diagonal fluctuation will have tachyonic mode.    After diagonalizing the tachyonic matrix field we see that the eigenfunction can describe the new recombined branes.   In the section IV we extend our method to discuss the general behavior of the recombination of  intersection D-branes which possessing arbitrary function forms.    We also present the physical reasons behind the mathematical process of diagonalizing the tachyonic matrix field.

    Note that in a recent paper Hashimoto and Nagaoka [14] had investigated the recombination of intersecting D-branes by using the super Yang-Mills theory which are low energy effective theories of D-branes.   The Yang-Mills field therein represents the dynamics field on the branes while the Higgs fields represent the locations of the D-branes.    Our investigations are within the framework of tachyon field theory, thus the  dynamics field on the branes are these fluctuating in the background of kink solution and the locations of the D-branes are at the zeros of tachyon field (see the next section).   

\section{Tachyon Effective Field Theory of One Brane System} 

    The two derivative truncation of boundary string field theory which embody the tachyon dynamics is described by the Lagrangian [6-10]

$$L_{D_p}= -\tau _p \left(\frac{1}{2}\partial^\mu T\partial_\mu T + 1\right) e^{-T^2/4}. \eqno{(2.1)}$$ 
where $\tau_p$ is the tension of the unstable $D_p$ brane.  The associated field equation is 

$$4 \partial_\mu^2  T + T (2- \partial_\mu T \partial^\mu T) = 0. \eqno{(2.2)}$$ 

There are some interesting solutions in the above field equation:

   (1) $T=0$:   This is an unstable vacuum which corresponds to the original $D_p$ brane. Its energy density is exactly equal to the $D_p$ brane tension.

   (2) $T=\pm\infty$:   This is the stable vacuum which  is thought of as the "closed string vacuum" with vanishing energy.    

   (3) $T=\sqrt{2} x$:    This is a stable one soliton solution which represents a kink solution with center at $x=0$.  This kink solution has tension $\tau_{p-1}=2\sqrt{2\pi}\tau_p$ which is  reasonably close to the actual value of $\tau_{p-1}=\sqrt{2}\pi\tau_p$.   The mass square for the fluctuation modes about this soliton has the equal spacing and the mass tower starts from a massless state.   There has no tachyonic mode and we can therefore identify this solution as a lower-dimensional stable BPS brane [3-6].

      (4) $T= a  x \pm b y $, if $a^2+b^2=1$:   As can be seen from (2.1) that  the energy density of this kink solution is maximum along the trajectory $a x \pm b y=0$.   Therefore this solution represents  an another stable one soliton solution which represents a kink solution with center located at $a x \pm b y=0$. In fact, this kink solution may be regarded as the solution 3 while in a new coordinate.   The intersection angle $\theta$ between the x-axis and this kink solution  is 
$$ \theta = tan^{-1}(\pm b/a).   \eqno{(2.3)}$$

It is a difficult work  to find another kink solution from the field equation (2.2).  However, it shall be noticed that the Lagrangian (2.1) is a truncated effective one in which the higher-derivative terms have been neglected.  Therefore, to make sense the finding kink solution we see that the solution can not be curved to much.   The solutions  3 and 4 are surely satisfy the criteria.   The fluctuation from such a kink solution is the same as the solution 3 and will be investigated in section 3.1.   It will be seen that there is no tachyonic mode as the kink-type condensation is a lower dimensional stable BPS brane.

\section{Tachyon Effective Field Theory of Two Branes System} 

   In order to describe two branes system the tachyon field needs to be generalized to a two-by-two matrix [8-10].   In this paper we use the following Lagrangian of the tachyon effective field   

   $$L=  -\tau_p Tr \left[~ \frac{1}{2} \partial^ \mu T e^{-T^2/8}\partial_\mu T e^{-T^2/8} + e^{-T^2/4}~\right].   \eqno{(3.1)}$$
This Lagrangian has been derived from the string theory by Kutasov, Marino and Moore [10].   Minahan [9] had also used this Lagrangian to investigate the property of stretched strings in tachyon condensation models.

  For the two-branes system the tachyonic matrix field can be expressed as 

$$T= \left(\begin{array}{cc} T_+ +t_+ & f \\f& T_- + t_- \end{array}\right)=\left(\begin{array}{cc} T_0+\triangle T_0 + t_0+\triangle t_0 & f \\f& T_0-\triangle T_0+t_0-\triangle t_0 \end{array}\right).\eqno{(3.2)} $$
\\
When $T_0= a x $, $\triangle T= b y$ with $a^2+b^2 = 2$ then the diagonal parts,  $T_\pm \equiv a x \pm b y $ represent kinks (i.e., a stable BPS branes after tachyon condensation from a higher dimensional unstable Non-PBS branes) with the centers located at $a x \pm b y = 0$. The another diagonal fields $t_\pm (x,y) \equiv t_0 \pm \triangle t_0$ describe  the fluctuations on the  kink solutions $T_\pm$.   The off-diagonal fields are $f(x,y)$  which represent the lowest energy excitation of the strings stretched between a BPS brane (i.e., a kink located at $a x + b y = 0$) and BPS brane (i.e., a kink located at $ a x - b y = 0$).

\subsection{Diagonal Fluctuation of Two Branes System} 

  Let us first neglect the off-diagonal part in the tachyon field matrix (3.2) and investigate the fluctuations of the diagonal fields $t_\pm (x,y)$.   In this case we can represent the tachyon field as

$$T= \left(\begin{array}{cc} \tilde T_0+ \triangle \tilde  T_0 & 0 \\0& \tilde T_0 - \triangle \tilde  T_0  \end{array}\right) = \tilde T_0 + \sigma_3 \triangle \tilde  T_0,   \eqno{(3.3)}$$
in which  we define $\tilde T_0 = T_0+t_0$,  $\triangle \tilde T_0 = \triangle T_0 + \triangle t_0$.   $\sigma_3$ is a Pauli matrix.   Then from the simple relations $[1,\sigma_3]=0$, $\sigma_3^2 =1$ and $e^{a \sigma_3}= cosh(a) + \sigma_3 sinh(a) $ we can find that 

$$e^{-T^2/8} = e^{-\left(\tilde T_0^2+\triangle \tilde T_0^2\right)/8}\left[cosh \left( \tilde T_0 \triangle \tilde T_0/4\right) - \sigma_3 ~ sinh \left( \tilde T_0  \triangle \tilde T_0 /4\right)\right],  \eqno{(3.4a)}$$
$$\partial _\mu T =\partial _\mu \tilde T_0  + \sigma_3 \partial _\mu \triangle \tilde T_0. \hspace{7cm} \eqno{(3.4b)}$$
Substituting above relations into (3.1) the Lagrangian expanded to the second order of the fluctuation fields $t_\pm$ can be found.   The associated field equations are  

$$4 \partial_\mu^2  T_\pm + T_\pm (2- \partial_\mu T_\pm\partial^\mu T_\pm) = 0, \eqno{(3.5)}$$ 
which is just the equation (3.2) and have the solutions  $T_\pm \equiv a x \pm b y $, which represent kinks with the centers located at $a x \pm b y = 0$, if $a^2+b^2 = 2$.   For such a system of intersecting branes the fluctuation field $t_\pm$ will satisfy the field equation

$$ 2 \left({\partial^2 t_\pm\over \partial x^2} + {\partial^2 t_\pm\over \partial y^2}\right) - (a x \pm b y) \left(a {\partial  t_\pm\over \partial x} \pm b {\partial t_\pm\over \partial y}\right) = m^2~ t_\pm , \eqno{(3.6)}$$
The mode function and mass square of the above equation are
$$ f_n(x,y) = (n!)^{-1/2} 2^{-n/2} \left({a^2\over 4\pi}\right)^{1/4} H_n(a x \pm b y), \eqno{(3.7)}$$
$$ m^2 = {1\over 4} n, ~~~~n\geq 0. \hspace{3cm}  \eqno{(3.8)}$$
The above solution describes a simple harmonic oscillation and the mass tower starts from a massless state,  thus there has no tachyonic mode in there.   These solutions are just the solution 3 or 4 in the section 2.    This is a trivial phenomena as neglecting the off-diagonal interaction the two branes become independent.    

\subsection{Off-diagonal Fluctuation of Two Branes System} 

   Let us now neglect the diagonal fluctuation field $t_\pm$ part in the tachyon field matrix (3.2) and investigate the fluctuations of the off-diagonal fields $f(x,y)$.   In this case we can represent the tachyon field as

$$T= \left(\begin{array}{cc} T_+  & f \\f& T_- \end{array}\right) = \left(\begin{array}{cc} T_0+\triangle T_0 & f \\f& T_0-\triangle T_0\end{array}\right) = T_0 + \sigma_3 \triangle T_0 + \sigma_x f.   \eqno{(3.9)} $$
Then, using the formula

$$e^{x \sigma_x + z \sigma_z}=cosh(\sqrt{x^2+z^2})  + {x\sigma_x+z\sigma_z\over \sqrt{x^2+z^2}} sinh(\sqrt{x^2+z^2}),   \eqno{(3.10)}$$
the Lagrangian (3.1) expanding to the second order in the off-diagonal tachyon field can be found.   When $T_\pm = ax \pm bz$, which represent intersecting kinks with the centers located at $a x \pm b y = 0$, if $a^2+b^2 = 2$, then the field $f(x,y)$, which represents the lowest string mode connecting the intersecting kinks, will satisfy the equation

$$ \left({\partial^2 f\over \partial x^2} + {\partial^2 f\over \partial y^2}\right) - {1\over 2} \left( a^2 x~{\partial  f\over \partial x} + b^2 y~{\partial f\over \partial y}\right) - {1\over 4} (a^2-b^2-2) f= m^2~f, \eqno{(3.11)}$$
\\
The mode solution and mass square of the above equation are
$$ f_{k,n}(x,y) =\phi_k(x) \phi_n(y), \hspace{6cm}\eqno{(3.12)}$$
$$\phi_k(x) \equiv (k!)^{-1/2} 2^{-k/2} \left({a^2\over 4\pi}\right)^{1/4} H_k(a x), $$
$$\phi_n(y) \equiv (n!)^{-1/2} 2^{-n/2} \left({b^2\over 4\pi}\right)^{1/4} H_n(b y), $$
$$ m^2 = {1\over 2} (a^2 k + b^2 (n-1)) , ~~~~k, n\geq 0. \hspace{3.5cm}  \eqno{(3.13)}$$
\\
The above solutions describes simple harmonic oscillations and the mass tower starts from a negative state with $m^2 = - {b^2\over2}$ which corresponds to the mode of $k=n=0$.  Therefore there has tachyonic mode in the intersection D-branes when the interaction therein is turned on.  The existence of the tachyonic mode  signs the instability of the intersection D-branes.   The tachyonic mode in here does not depend on the space coordinate.    This property was first found in [6].

    Note that the worldsheet spectrum of the string connecting the intersecting D-branes has a tachyonic state whose mass squared is $m^2= |b|$ for small $b$ [12].   This contradicts our results.   In fact, our tachyonic field system would not be changed if $b$ is replaced by $-b$.   Thus, as it seems difficult to obtain a spectrum proportional to $|b|$ in solving the tachyonic field equation, the spectrum proportional to $b^2$ seem to be a general result if one try to study the problem from the tachyonic field model.   We conjecture that such a discrepancy may be a general property in a tachyon effective field theory even if the higher-derivative terms were included \footnote{After the paper was released to the arXiv Hashimoto and Nagaoka [14] informed us that they had also known such a discrepancy.}.

    It was argued that [11-14] a pair of  intersection D-branes will recombine.    To describe the new recombined state we will consider the following tachyonic matrix field

$$T_{initial}= \left(\begin{array}{cc} T_+  & f_{0,0} \\f_{0,0}& T_- \end{array}\right) = \left(\begin{array}{cc} a x+ b y & {\sqrt {a b\over 4}} \\{\sqrt {a b\over 4}}& a x - b y \end{array}\right),   \eqno{(3.14)} $$
in which $f_{0,0}$ is the tachyonic mode.   Then, after diagonalizing the above two-by-two tachyonic matrix field we see that the eigenfunctions describe the new recombined branes are

    $$T_{new} = a x \pm \sqrt {b^2 y^2 + {ab\over 4\pi}}. \eqno{(3.15)} $$


The geometry of the recombination is shown in figure 1.
\\

\hspace{6cm}\scalebox{0.7}{\includegraphics{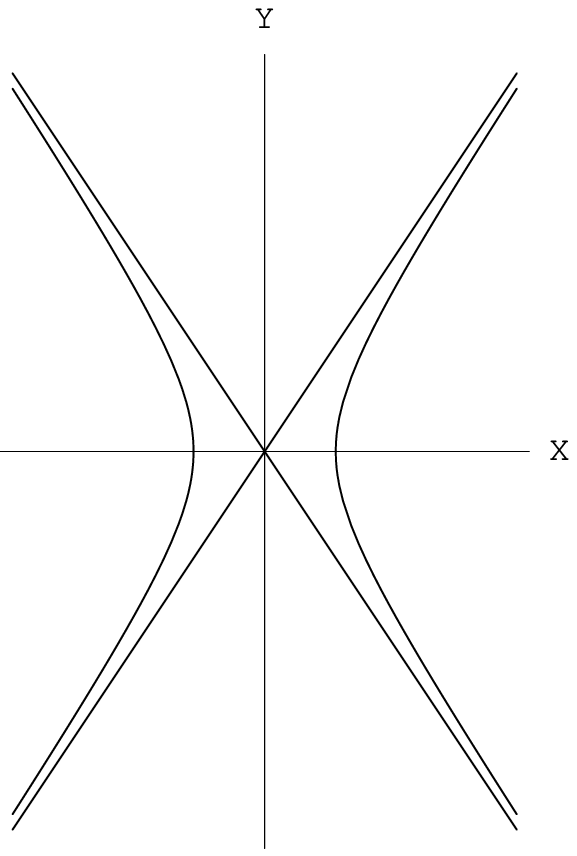}} 
\\
\\
{\it ~~~Fig.1. Recombination of intersecting branes.   The original pair of branes are intersecting at (0,0).  After the recombination the new branes, which do not intersect to each other, will cross x-axis near the positions $(\sqrt {{ab\over 4\pi}},0)$ and ($- \sqrt {{ab\over 4\pi}},0)$ respectively. }
\\

    Note that Hashimoto and Nagaoka [14] investigated the recombination of Intersecting D-branes by using the super Yang-Mills theory which are low energy effective theories of D-branes.   In their approach the Higgs fields therein represent the {\it locations} of the D-branes.   They diagonalize the matrix, which is constructed from  the original Higgs fields and the associated small fluctuation, to find the new eigenfunctions which, they claim to be able to represent the new {\it locations} of the combined branes.       Our approach, on the other hand, use the effective tachyon field theory.   We regard the branes as the  kink-type tachyon condensed states and use the effective two-tachyon Lagrangian to study the off-diagonal fluctuation therein.   We describe the new eigenfunctions as the configurations of the recombined branes.  Our method give an alternative view to see  the recombination of intersecting branes.   In the next section we will use our prescription to discuss the general behavior of the recombination of  intersection D-branes which possessing arbitrary function forms.   We also present the physical reasons behind the mathematical process of diagonalizing the tachyonic matrix field.

\section{General property of a Pair of Intersecting  Branes }
Consider the more general case in which the lower-dimensional stable BPS D-brane solutions are described be the tachyon fields $T_\pm(x,y)$.  Due to the potential form $e^{-T^2/8}$ in (3.1) the branes is located near the position $T_\pm(x,y) \approx 0$.    If the branes are intersecting then we can choose a new coordinate such that near the intersecting position the branes recombination can also be described as Figure 1.  Therefore the mechanism of the recombination near the intersecting position shall be just that described in section 3.    The other parts of the brane far away from the intersecting position does not be changed significantly as can be seen from the following argument.

  For intersecting branes described by the tachyon fields $T_\pm(x,y)$ there will have a tachyonic mode in the off-diagonal fluctuation $f_0(x,y)$, which signs the instability of the intersecting D-branes.  The generally tachyonic matrix field can be expressed as 
$$T_{initial}=\left(\begin{array}{cc}T_+(x,y)& f_0(x,y)\\f_0(x,y)& T_- (x,y) \end{array} \right) .   \eqno{(4.1)} $$
in which $f_0(x,y)$ is a small fluctuation of the tachyonic mode.  Now, after diagonalizing the above two-by-two tachyonic matrix field we see that the eigenfunctions describe the new recombined branes are
    $$T_{new} = {1\over2}\left(T_++T_ -\pm \sqrt {(T_+-T_- )^2+ f_0^2}\right). \eqno{(4.2)}$$
We see that $T_{new} \not= 0$ at the initial intersection positions   in which  $T_+(x,y)=T_-(x,y)=0$. Thus the intersecting branes shall be recombined.   The positions satisfy the relations  $T_\pm(x,y)=0, T_\mp (x,y)\not=0$ represent a brane  described by tachyon field $T_\mp (x,y)$ while excluding the intersecting positions.  In this case we see that, as the fluctuation field is small,  $T_{new} \approx T_\mp (x,y)$. Thus the recombined brane far away from the intersecting positions is approximately described by the original tachyon field $T_\mp (x,y)$.  We can therefore conclude that the parts of the brane far away from the intersecting position does not be changed significantly.

    We finally present the physical reasons behind the mathematical process of diagonalizing the tachyonic matrix field.

   (1) Initially we prepare a pair of intersecting D-branes with kink fields $T_\mp$.   After turning on the interaction between the intersecting D-branes there will appear tachyonic mode fluctuation $f_0$ which signs the instability of the system.   The tachyonic matrix field $T_{initial}$ with diagonal part $T_\mp$ and off-diagonal part $f_0$ was used to represent the initially unstable state.

  (2)  To stabilize the system the unstable intersecting D-branes, which is represented by $T_{initial}$, shall be recombined into two separated D-branes which would never intersect to each other anymore, because any intersecting D-branes will possess tachyonic mode fluctuation and is thus still unstable.   This means that the kink fields $T_\mp$ shall combine with tachyon field $f$ and  condense into two separated D-branes.    In this case we can therefore use a diagonalized new tachyonic matrix field $T_{new}$ to represent the new stable state. 

   (3) Now, from the {\it energy conservation} we know that the system described by the tachyonic matrix fields $T_{initial}$ and $T_{new}$ shall have the same energy.   Therefore, denoting the Hamiltonian as H, we have a relation 
            $$Tr~H(T_{initial}) = Tr~H(T_{new}), \eqno{(4.3)}$$
(Note that as the initial and new systems are both described by the same Lagrangian $L$ (3.1) they shall have the same Hamiltonian form $H$ too.) 

  (4) One can then try to find a matrix $S$ to diagonalize the $T_{initial}$,
             $$ T_{new} = ST_{initial}S^{-1}, \eqno{(4.4)}$$    
and substitute (4.4) into (4.3) to check whether the relation (4.3) was satisfied.   If it was satisfied then the new recombined brane is possible to be represented by the new matrix tachyon field $T_{new}$.   Note that any theory has many states which have the same energy so the checking that the energies are identical could not sure that the states are the same.   Therefore the above argument is just only one of  the supporting evidences. 

    In short, if the matrix field $T_{initial}$, which represents initially unstable intersecting D-branes, and {\it its diagonalized matrix field} $T_{new}$, which represents two separated stable D-branes, have the same energy then the matrix field $T_{new}$ may be able to represent the new recombined D-branes.

   It shall be noted that as there is the term $\partial_{\mu}T$ in $H$ it is not easy to perform such a checking work.   However, when the fluctuation field is small then the checking work becomes more easy, as the matrix $S$ is a unit matrix plus a matrix with small elements.   Our checking has found that the relation (4.3) is consistent with (4.4) and we thus claim that $T_{new}$ can really represent  the new recombined D-branes.   

\section {Conclusion}
   In this paper we investigate the mechanism of recombination of intersection D-branes within the framework of the effective tachyon field theory.   We use the effective two-tachyon matrix field Lagrangian to study the off-diagonal fluctuation in the background of the kink solutions.   We regard the kink solutions as the tachyon condensations of the non-BPS brane in higher dimension.     In the background of the  intersecting D-branes the off-diagonal fluctuations have tachyonic mode.    After diagonalizing the tachyon matrix field we see that the eigenfunction can describe the new recombined branes.   We generalize our method to discuss the general behavior of the recombination of  intersection D-branes which possessing arbitrary function form.    We also present in detail the physical reasons behind the mathematical process of diagonalizing the tachyonic matrix field.   It is hoped the our prescription can be extended to discuss other unstable brane configurations such as the brane ending on branes, brane system with different dimensions, etc.   We will investigate these problems in the later studying.

\newpage

\begin{enumerate}
\item  J. Polchinski, ``String Theory'', Cambridge University Press,
1998.
\item W. Taylor, ``M(artrix) Theory: Matrix Quantum Mechanics as a Fundamental Theory'',  Rev. Mod. Phys. 73 (2001) 419, hep-th/0101126.
\item  A. Sen,  ``Non-BPS States and Branes in String Theory'',  hep-th/9904207.
\item  A. Lerda and R. Russo, ``Stable Non-BPS States in String Theory'',  Int.\ J.  Mod.  Phys.  A15 (2000) 771, hep-th/9905006.
\item A. Sen,  ``Tachyon Condensation on the Brane Antibrane System'', 
  JHEP 9808  (1998)  012,   hep-th/9805170;   ``Descent Relations Among Bosonic D-branes'',  Int.\ J.\ Mod.\ Phys.  A14 (1999) 4061,  hep-th/9902105;   ``Universality of the Tachyon Potential'',  JHEP  9912 (1999) 027, hep-th/9911116.
\item S.\ Moriyama and S.\ Nakamura,   ``Descent Relation of Tachyon Condensation from Boundary String  Field Theory'' ; Phys. Lett.  B506 (2001) 161;  hep-th/0009246 ; K. Hashimoto and S. Nagaoka,  ``Realization of Brane Descent Relations in Effective Theories'' , Phys. Rev  D66  (2002) 02060011,   hep-th/0202079 . 
\item  B.\ Zwiebach,   ``A Solvable Toy Model for Tachyon Condensation in String Field  Theory'' ,   JHEP  0009  2000  028 ,   hep-th/0008227 ; J.\ A.\ Minahan and B.\ Zwiebach,   ``Field Theory Models for Tachyon and Gauge Field String  Dynamics'' ,  JHEP  0009  2000  029 ,   hep-th/0008231 .  J.\ A.\ Minahan and B.\ Zwiebach,   ``Effective Tachyon Dynamics in Superstring Theory'' ,    JHEP  0103  2001  038 ,    hep-th/0009246 .  J.\ A.\ Minahan and B.\ Zwiebach,    ``Gauge Fields and Fermions in Tachyon Effective Field  Theories'' , JHEP  0102  2001  034 ,   hep-th/0011226.
\item  A.~A.~Gerasimov and S.~L.~Shatashvili, ``On non-abelian structures in field theory of open strings,'' JHEP {\bf 0106}, 066 (2001) [arXiv:hep-th/0105245]; E.~T.~Akhmedov, ``Non-Abelian structures in BSFT and RR couplings'',  hep-th/0110002. V.~Pestun, ``On non-Abelian low energy effective action for D-branes'', JHEP {\bf 0111}, 017 (2001), hep-th/0110092.
\item   Minahan,   ``Stretched Strings in Tachyon Condensation Models'' ,  JHEP  0205  2002  024 , hep-th/0203108.
\item    D. Kutasov, M.\ Marino and G.\ Moore,  ``Remarks on Tachyon Condensation in Superstring Field  Theory'', hep-th/0010108. 
\item D. J. Smith, ``Intersecting Brane Solution in string and M-Theory'',  hep-th/0210157. 
\item  M.~Berkooz, M.~R.~Douglas and R.~G.~Leigh, ``Branes Intersecting at Angles'', Nucl. Phys. B480 (1996) 265, hep-th/9606139; M. M. Sheikh Jabbari,``Classification of Different Branes at Angles'',  Phys. Lett. B420 (1998) 279, hep-th/9710121; N. Ohta and P. Townsend,``Supersymmetry of M-Branes at Angles'',  Phys. Lett. B418 (1998) 77, hep-th/9710129.
\item A.~Hashimoto and W.~Taylor IV, ``Fluctuation Spectra of Tilted and Intersecting D-branes from the Born-Infeld Action,'' Nucl. Phys. B503 (1997) 193, hep-th/9703217. 
\item K.~Hashimoto and S.~Nagaoka, ``Recombination of Intersecting D-branes by Local Tachyon Condensation'',  hep-th/0303204. 
\item R.~Blumenhagen, B.~K\"ors, D.~L\"ust and T.~Ott,  ``The Standard Model from Stable Intersecting Brane World Orbifolds'', Nucl. Phys. B616 (2001) 3, hep-th/0107138; D.~Cremades, L.~E.~Ibanez and F.~Marchesano,
``Intersecting Brane Models of Particle Physics and the Higgs Mechanism'', 
JHEP 0207 (2002) 022, hep-th/0203160; C.~Kokorelis, ``Exact Standard model Structures from Intersecting D5-Branes'', hep-th/0207234.
\item J.~Garcia-Bellido, R.~Rabadan and F.~Zamora, ``Inflationary Scenarios from Branes at Angles'', JHEP 0201 (2002) 036, hep-th/0112147; R.~Blumenhagen, B.~K\"ors, D.~L\"ust and T.~Ott, ``Hybrid Inflation in Intersecting Brane Worlds'', Nucl. Phys. B641 (2002) 235, hep-th/0202124;
N.~Jones, H.~Stoica and S.~-H.~Henry Tye, ``Brane Interaction as the Origin of Inflation'',  JHEP 0207 (2002) 051,  hep-th/0203163; S.~Sarangi and S.~-H.~Henry Tye, ``Cosmic String Production Towards the End of Brane Inflation'', Phys. Lett. B536 (2002) 185,  hep-th/0204074; M.~Gomez-Reino and I.~Zavala,
``Recombination of Intersecting D-Branes and Cosmological Inflation'', JHEP 0209 (2002) 020, hep-th/0207278.
\end{enumerate}
\end{document}